\documentclass[pra,twocolumn,a4paper]{revtex4}
\usepackage{graphicx}
\usepackage{amsmath}
\usepackage{bm}
\renewcommand{\Re}{\operatorname{Re}}
\renewcommand{\Im}{\operatorname{Im}}
\DeclareMathOperator{\sgn}{sgn}

\begin{document}

\title{A Photon-Photon Quantum Gate Based on Rydberg Interactions}

\author{Daniel Tiarks}
\author{Steffen Schmidt-Eberle}
\author{Thomas Stolz}
\author{Gerhard Rempe}
\author{Stephan D\"{u}rr}
\affiliation{Max-Planck-Institut f\"{u}r Quantenoptik, Hans-Kopfermann-Stra{\ss}e 1, 85748 Garching, Germany}

\maketitle

{\bf
The interaction between Rydberg states of neutral atoms is strong and long-range, making it appealing to put it to use in the context of quantum technologies. Recently, first applications of this idea have been reported in the fields of quantum computation \cite{Saffman:16} and quantum simulation \cite{Schauss:15,Labuhn:16,Bernien:17}. Furthermore, electromagnetically induced transparency allows to map these Rydberg interactions to light \cite{Lukin:01a,Friedler:05,Gorshkov:11,Pritchard:10,Firstenberg:13,Baur:14,Gorniaczyk:14,Tiarks:14,Tiarks:16,Ningyuan:16,Thompson:17}. Here we exploit this mapping and the resulting interaction between photons to realize a photon-photon quantum gate \cite{O'Brien:03,Hacker:16}, demonstrating the potential of Rydberg systems as a platform also for quantum communication and quantum networking \cite{Kimble:08}. We measure a controlled-NOT truth table with a fidelity of 70(8)\% and an entangling-gate fidelity of 63.7(4.5)\%, both post-selected upon detection of a control and a target photon. The level of control reached here is an encouraging step towards exploring novel many-body states of photons or for future applications in quantum communication and quantum networking \cite{Kimble:08}.
}

Optical technologies serve as today's standard for distributing information in the Internet since photons offer high speed and large bandwidth. Because of these benefits, future quantum technologies will probably rely on photonic qubits to transfer quantum states between distant nodes. However, ambitions to use photons for processing rather than only transmitting qubits are hampered by the fact that photons hardly interact with each other. A solution to this problem is offered by a Rydberg polariton \cite{Friedler:05,Gorshkov:11,Pritchard:10,Firstenberg:13,Baur:14,Gorniaczyk:14,Tiarks:14,Tiarks:16,Ningyuan:16,Thompson:17}. This intriguing quasiparticle - composed of a photonic component and an atomic Rydberg excitation - is obtained when a photon enters a medium in which electromagnetically induced transparency (EIT) couples the photon to a Rydberg state. The key idea is that the strong, long-range interaction between the atomic Rydberg components is mapped onto the photons. The potential for applying this to build a photon-photon gate as been discussed at great length in the literature. The importance of this goal can be illustrated by the large number of schemes proposed for building a Rydberg-based photon-photon quantum gate \cite{Friedler:05,Gorshkov:11,He:14,Paredes-Barato:14,Khazali:15,Hao:15,Das:16,Wade:16,Murray:17,Lahad:17}. In addition, the demonstration of a photon-photon quantum gate is the hallmark for having achieved full quantum control of one photon over another.

\begin{figure}[!t]
\centering
\includegraphics[width=\columnwidth]{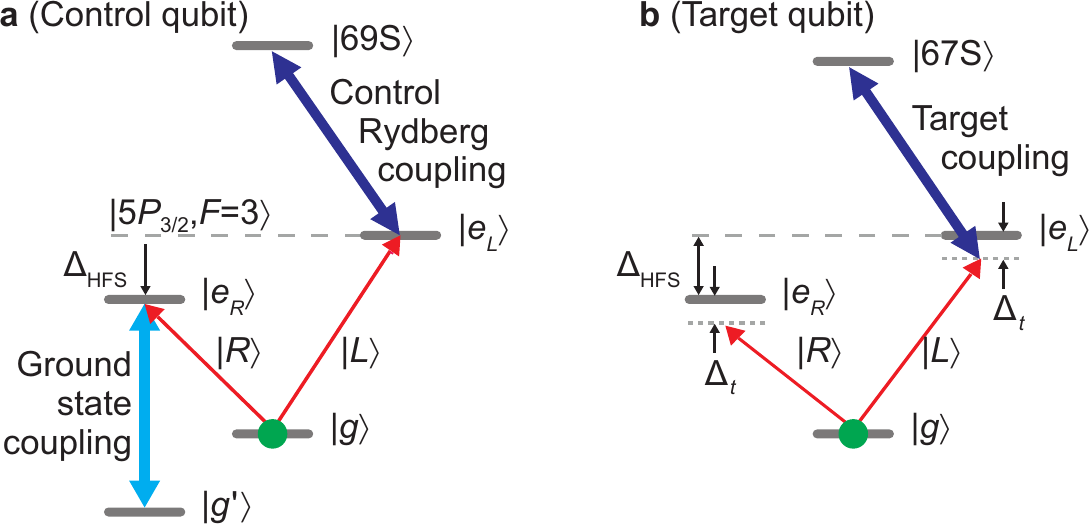}
\caption{
Atomic level schemes. \textbf{a}, Level scheme for EIT storage of the control qubit. The initial population (green) is prepared in state $|g\rangle$. $|L\rangle$ polarisation is stored in a Rydberg state $|69S\rangle$, $|R\rangle$ polarisation in a ground state $|g'\rangle$. \textbf{b}, Level scheme for the target qubit. $|L\rangle$ polarisation propagates as a Rydberg polariton involving Rydberg state $|67S\rangle$, $|R\rangle$ polarisation propagates off-resonantly without experiencing EIT.
}
\label{fig-level-scheme}
\end{figure}

Here we report on the first experimental demonstration of a Rydberg-based photon-photon quantum gate. An important ingredient is the creation of a conditional $\pi$ phase shift. To achieve this, first, a control photon is stored in an ultracold atomic ensemble based on EIT, in such a way that the left- and righthanded circular polarisations $|L\rangle$ and $|R\rangle$ are stored in a Rydberg state $|69S\rangle$ and a ground state $|g'\rangle$, respectively (Fig.\ \ref{fig-level-scheme} and Methods). Second, a target photon propagates through the ensemble. Its $|L\rangle$ polarisation is coupled to another Rydberg state $|67S\rangle$ by EIT, making it a Rydberg polariton, while its $|R\rangle$ polarisation simply propagates off-resonantly through the ensemble. Only if both photons have $|L\rangle$ polarisation, will they both have a Rydberg component, causing them to interact. This interaction modifies the linear electric susceptibility $\chi$ experienced by the target photon. And that modifies the phase shift which the target photon accumulates during propagation through the ensemble. For appropriate parameters the resulting conditional phase shift equals $\pi$. Finally, the control photon is retrieved. This realizes a controlled phase flip gate. A change of basis easily converts this into a controlled-NOT (CNOT) gate.

Each incoming qubit is implemented as an attenuated pulse of laser light with Poissonian photon number statistics and mean photon number below unity. Data are post-selected upon detection of one photon in each pulse to compensate for non-unity efficiencies and for the photon statistics of the input. The efficiency of the gate, i.e.\ the probability that the atomic ensemble transmits a photon pair impinging onto it, ranges between 8\% and 0.5\% depending on input polarisations.

In our experiment, the conditional $\pi$ phase shift is accumulated from the interaction of the two wave packets during their spatiotemporal overlap. This is a long-standing goal since the 1980s, where it was the basis for the first proposal for a quantum gate for photons \cite{Milburn:89}. Previous experimental realisations of photon-photon quantum gates circumvented the difficulty in creating an interaction between overlapping wave packets. As a workaround, they relied either on linear optics combined with inherently probabilistic protocols \cite{O'Brien:03} or on a sequence of two atom-photon gates \cite{Hacker:16}.

We previously demonstrated one important ingredient of the gate, namely a conditional phase shift $\Delta\beta$ of $\pi$ \cite{Tiarks:16}. The present experiment features two crucial advances that make it possible to realize a quantum gate. First, achieving high post-selected fidelity requires that the conditional optical depth $\Delta OD$ vanishes because the ideal gate has vanishing $\Delta OD$ and single-qubit operations cannot change $\Delta OD$ (Methods). When trying to simultaneously reach $\Delta\beta= \pi$ and $\Delta OD= 0$, the main limiting factor is dephasing, i.e.\ decay of coherence between the ground state $|g\rangle$ and the Rydberg state $|67S\rangle$. This makes it nontrivial to find suitable experimental parameters (Methods).

Second, in Ref.\ \cite{Tiarks:16} we conditioned the $\pi$ phase shift on the presence or absence of a control photon to be stored in state $|69S\rangle$. The presence or absence of a photon represents a qubit, but using the qubit in this form is inexpedient in an experiment because photon loss will cause bit-flip errors and because single-qubit unitaries are difficult to realize. We solve this problem by mapping this qubit onto a polarisation qubit so that photon loss can be handled by post-selection and single-qubit unitaries are easy to implement using wave plates. The $|L\rangle$ polarisation of the qubit is stored in state $|69S\rangle$. As the $|L\rangle$ and $|R\rangle$ polarisations of this qubit must interfere at the detector, the $|R\rangle$ polarisation must be delayed as much as the stored $|L\rangle$ polarisation. In principle, this could be achieved in several ways (Methods). We choose to store the $|R\rangle$ polarisation in the $|g'\rangle= |5S_{1/2},F{=}m_F{=}1\rangle$ ground state, where $F,m_F$ are the hyperfine quantum numbers.

\begin{figure}[!b]
\centering
\includegraphics[width=\columnwidth]{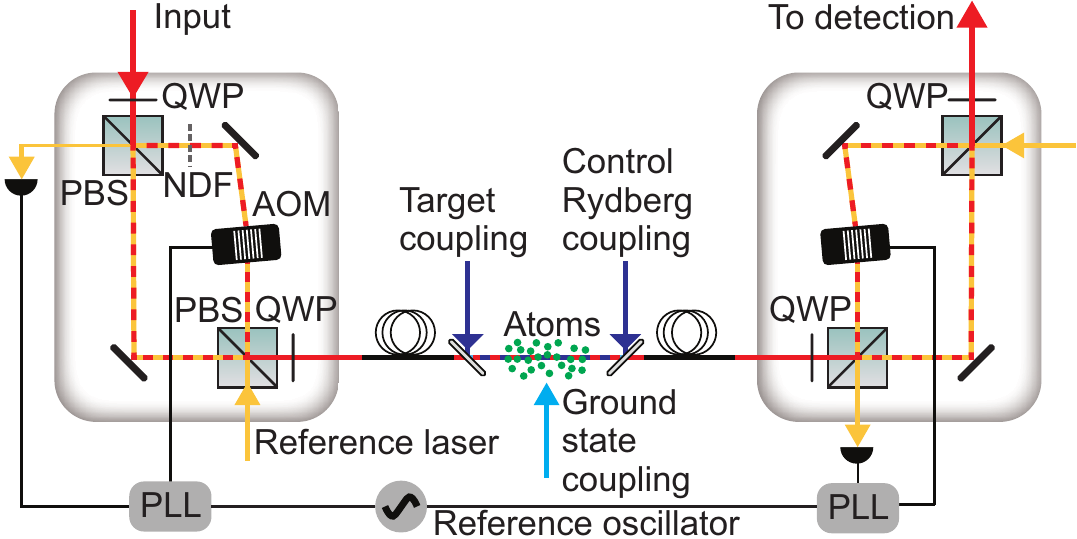}
\caption{
Simplified scheme of the experimental setup. The control photon (red) passes through a first interferometer (left box), interacts with the atomic ensemble (green), and finally passes through a second interferometer (right box). The first interferometer shifts the frequency of only the $|R\rangle$ polarisation by $\Delta_\mathrm{HFS}/2\pi$. The second interferometer removes this frequency shift. Counterpropagating reference light (yellow) and phase-locked loops (PLLs) stabilise each interferometer against thermal drift. Coupling light (blue, cyan) establishes EIT. The target photon takes the same path as the control photon.
}
\label{fig-setup}
\end{figure}

For reasons discussed in the Methods section, we choose the frequency of the incoming control photon to be resonant with the $|g\rangle \leftrightarrow |5P_{3/2},F{=}3\rangle$ transition (Fig.\ \ref{fig-level-scheme}a). As a result, achieving storage in state $|g'\rangle$ is not straightforward because the states $|g'\rangle$ and $|5P_{3/2},F{=}3\rangle$ are not connected by an electric-dipole transition because of $\Delta F=2$. We overcome this hurdle by frequency shifting selectively only the $|R\rangle$ component of the control photon to make it resonant with the transition from $|g\rangle$ to $|e_R\rangle= |5P_{3/2},F{=}2,m_F{=}1\rangle$. From here, we achieve resonant EIT-based storage in $|g'\rangle$. To create the required frequency shift of $\Delta_\mathrm{HFS}/2\pi= 267$ MHz, we use an acousto-optic modulator (AOM). To shift the frequency of only one polarisation, we first spatially split the polarisations with a quarter-wave plate (QWP) and a polarising beam splitter (PBS), send only one of the beams through the AOM, and finally recombine the beams using another PBS and QWP. Overall, this realizes a modified Mach-Zehnder interferometer which creates a frequency shift for only $|R\rangle$ polarisation (left half of Fig.\ \ref{fig-setup}).

The storage is achieved by simultaneously switching off both control coupling light fields while the control photon is inside the medium. After a dark time set to 4.5 $\mu$s, both control coupling light fields are simultaneously switched back on and both polarisation components of the control qubit are retrieved. To remove the frequency shift of the $|R\rangle$ component, the retrieved light is sent through a similar interferometer (right half of Fig.\ \ref{fig-setup}). Overall, this realises an EIT-based quantum memory, which is sophisticated insofar as this memory stores one polarisation component in a Rydberg state and the other in a ground state. This has the decisive advantage that only the Rydberg component will strongly interact with the target photon when it propagates through the medium during the dark time, thus making a quantum gate possible. The average post-selected fidelity \cite{Bowdrey:02} of the quantum memory is $F_m= 87.5(7)$\% (Appendix), which clearly exceeds the classical limit 2/3.

To build a photon-photon gate, we combine this quantum memory for the control qubit with the conditional $\pi$ phase shift of the target photon, as described above. On top of that, there are single-qubit phase shifts (Methods), which we choose such that the complete gate operation should ideally yield a truth table $|RR\rangle\mapsto|RR\rangle$, $|RL\rangle\mapsto-|RL\rangle$, $|LR\rangle\mapsto|LR\rangle$, and $|LL\rangle\mapsto|LL\rangle$, where the control qubit is listed first. This constitutes a controlled phase flip gate, which is a universal two-qubit gate.

\begin{figure}[!t]
\centering
\includegraphics[width=\columnwidth]{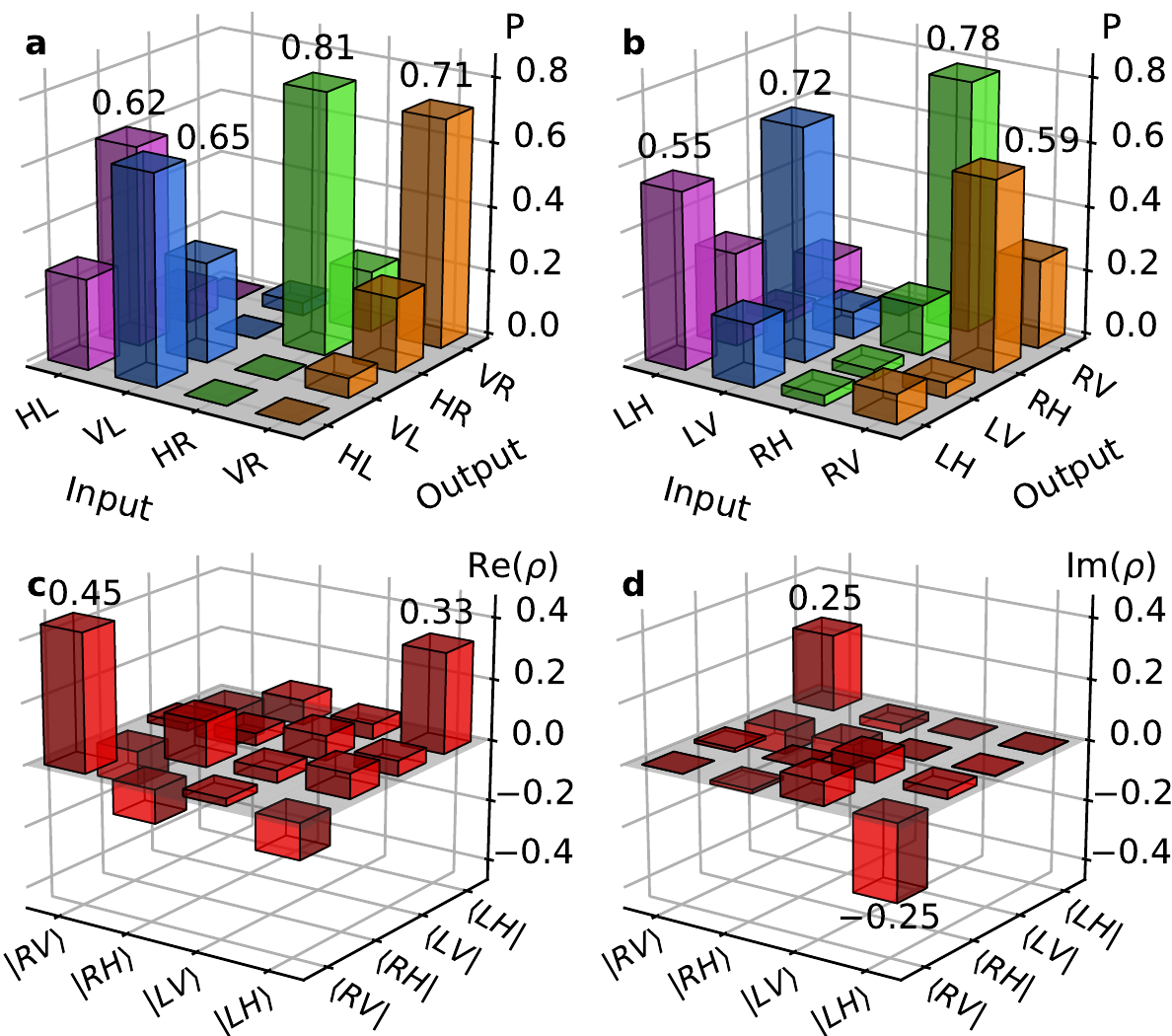}
\caption{
Performance of the photon-photon gate. \textbf{a}, Truth table of the CNOT gate operation. To obtain a truth table, one of the four displayed input states is prepared repeatedly and the probability $P$ to obtain the displayed output states is measured. Ideally, the polarisations $|H\rangle$ and $|V\rangle$ of the first qubit should be exchanged if and only if the second qubit is in $|L\rangle$. \textbf{b}, Same as in part a but with the input polarisations of the first and second qubit swapped. \textbf{c}, \textbf{d}, Real and imaginary parts of the reconstructed density matrix $\rho$ of the two-photon state created in the entangling-gate operation.
}
\label{fig-CNOT-entangle}
\end{figure}

We now turn to the characterisation of this gate. Fig.\ \ref{fig-CNOT-entangle}a shows a measured CNOT truth table in which the first photon is initially in the horizontal $|H\rangle= (|R\rangle+|L\rangle)/\sqrt2$ or vertical $|V\rangle= i(|R\rangle-|L\rangle)/\sqrt2$ polarisation state, while the second photon is initially in $|R\rangle$ or $|L\rangle$. The fidelity of this post-selected truth table, i.e.\ the probability of obtaining the desired output averaged over all four input states, is $F_\mathrm{CNOT}=70(8)$\%. Fig.\ \ref{fig-CNOT-entangle}b shows a similar CNOT truth table but with the input polarisations of the first and second photon swapped, yielding $F_\mathrm{CNOT}= 66(9)$\%. Studying the performance of both versions of the CNOT gate is interesting because the photons are treated differently insofar as one is stored and retrieved, whereas the other propagates without storage.

To demonstrate that the gate operates in the quantum regime, we study an entangling-gate operation. To this end, we prepare the input state $|HH\rangle$. From this, the gate should ideally produce the maximally-entangled output state $|\psi_i\rangle= (|LH\rangle-i|RV\rangle)/\sqrt2$. The actual output state can be described by a density matrix $\rho$. We measure a post-selected fidelity of $F_e=\langle\psi_i|\rho|\psi_i\rangle= 63.7(4.5)$\% (linear unbiased estimator). This is well above the threshold of 1/2 for demonstrating entanglement. Using quantum state tomography, we reconstruct the density matrix $\rho$ (linear unbiased estimator) (Fig.\ \ref{fig-CNOT-entangle}c,d). We note that in a simple model (Appendix), the non-unity visibilities of the control qubit and the target qubit already set an upper bound of 76(6)\% to $F_e$. The difference between the measured value and this upper bound indicates that there are additional imperfections in the gate operation beyond the single-qubit visibilities (Appendix).

The above results demonstrate the first realization of a Rydberg-based photon-photon quantum gate. It is a clear advantage of the gate scheme used here that its implementation requires manageable experimental effort. However, when aiming at high fidelity and high efficiency, other schemes \cite{Friedler:05,Gorshkov:11,He:14,Paredes-Barato:14,Khazali:15,Hao:15,Das:16,Wade:16,Murray:17,Lahad:17} might be better suited while an excitation-hopping based approach \cite{Thompson:17} presently seems to yield similar efficiency (Appendix). Nevertheless, we believe that efficiency and fidelity of the present scheme can both be improved. For example, EIT storage and retrieval with an efficiency of 92\% has been achieved recently \cite{Hsiao:18}. Future improvements in efficiency and fidelity might make high-efficiency photonic Bell state detection possible, which is crucial for quantum repeaters. Once efficiency and fidelity are improved, the present scheme could be extended to the generation of entangled states of several photonic qubits, by transmitting more than one target photon in the presence of the same stored control photon. Additionally, Rydberg-based quantum gates might allow for miniaturisation and transfer to solid-states systems. For example, interactions between Rydberg excitons in a solid have been observed recently \cite{Kazimierczuk:14}.

\begin{center}
{\large \textbf{Methods}}

\end{center}

\noindent
\textbf{Atomic Transitions}

\noindent
The experiment begins with the preparation of a gas of typically 10$^5$ ultracold $^{87}$Rb atoms at a temperature of typically 0.6 $\mu$K in an optical dipole trap (Appendix). The internal state is prepared in the $|g\rangle= |5S_{1/2},F{=}m_F{=}2\rangle$ ground state.

The incoming control photon is resonant with the $|g\rangle \leftrightarrow |5P_{3/2},F{=}3\rangle$ transition at a wavelength of 780 nm. The $|L\rangle$ polarisation of this photon is stored in the Rydberg state $|69S\rangle= |69S_{1/2},F{=}m_F{=}2\rangle$ based on resonant Rydberg EIT using control Rydberg coupling light at 480 nm and the intermediate state $|e_L\rangle= |5P_{3/2},F{=}m_F{=}3\rangle$. The $|R\rangle$ polarisation of the control photon is stored in the ground state $|g'\rangle= |5S_{1/2},F{=}m_F{=}1\rangle$ based on resonant EIT using ground-state coupling light at 780 nm and the intermediate state $|e_R\rangle= |5P_{3/2},F{=}2,m_F{=}1\rangle$ (Fig.\ \ref{fig-level-scheme}a and Appendix). To make the latter possible, the $|R\rangle$ polarisation of the control photon is frequency shifted using the modified Mach-Zehnder interferometer discussed above. In addition, this frequency shift creates a large detuning from the Rydberg-EIT resonance, thus preventing undesired storage of the $|R\rangle$ polarisation in a Rydberg state.

The incoming target photon is detuned by $\Delta_t/2\pi= -17$ MHz (to the red) from the $|g\rangle\leftrightarrow|e_L\rangle$ resonance. The $|L\rangle$ polarisation of this photon propagates through the medium as a Rydberg polariton, experiencing Rydberg EIT with intermediate state $|e_L\rangle$ (Fig.\ \ref{fig-level-scheme}b) and with Rydberg state $|67S\rangle= |67S_{1/2},F{=}m_F{=}2\rangle$, which features a F\"orster resonance with the $|69S\rangle$ state \cite{Tiarks:14}.
The target photon takes the same beam path through the setup as the control photon. In particular, it also passes through both interferometers. Hence, the $|R\rangle$ polarisation of the target photon, which enters the setup at the same frequency as the $|L\rangle$ polarisation, is $-\Delta_\mathrm{HFS}/2\pi= -267$ MHz detuned from the two-photon resonance for Rydberg EIT when reaching the atoms. This means that the $|R\rangle$ polarisation does not experience EIT. Instead, its dominant interaction with the medium comes from the $|g\rangle \leftrightarrow |e_R\rangle$ transition, from which it is detuned by $-17$ MHz.

\bigskip
\noindent
\textbf{Phase Shifts and Photon Loss}

\noindent
The combination of propagation, storage, and retrieval changes the polarisation state $|j\rangle$ of the control photon with $j\in\{R,L\}$ to $e^{-OD_{c,j}/2+i\beta_{c,j}}|j\rangle$ where $\beta_{c,j}$ is the accumulated phase shift and $OD_{c,j}$ expresses photon loss. We choose to operate the incoming control photon and both control coupling lasers at the single-photon resonances because this should minimize loss of the control photon. Theoretically, this should automatically yield $\beta_{c,L}-\beta_{c,R}= 0$. In practice, we fine tune the frequency of one of the control coupling lasers to achieve $\beta_{c,L}-\beta_{c,R}= 0$ for all dark times.

Propagation through the medium in the absence of a control qubit changes the polarisation state $|j\rangle$ of the target photon with $j\in\{R,L\}$ to $e^{-OD_j/2+i\beta_j}|j\rangle$ where $\beta_j$ is the accumulated phase shift and $OD_j$ the optical depth. The $|g\rangle \leftrightarrow |e_R\rangle$ transition has a relatively small electric dipole matrix element, making its resonant optical depth a factor of 6 smaller than for the $|g\rangle\leftrightarrow|e_L\rangle$ transition. Hence, the polarisation $|R\rangle$ experiences relatively high transmission $e^{-OD_R}\sim 0.77$ and a phase shift $\beta_R\sim 0.9$ rad, which is of little relevance.

If a control qubit is stored in Rydberg state $|69S\rangle$, then Rydberg blockade modifies the linear electric susceptibility $\chi$ experienced by the $|L\rangle$ polarisation of the target qubit. This changes the values of $OD_L$ and $\beta_L$ to $OD_{LL}$ and $\beta_{LL}$, respectively. The conditional optical depth is $\Delta OD= OD_{LL}-OD_L$ and the conditional phase shift $\Delta\beta= \beta_{LL}-\beta_L$. The control qubit affects neither $OD_R$ nor $\beta_R$. While $\Delta_t=0$ would be a good choice if one aimed for large $\Delta OD$ combined with $\Delta\beta=0$, we choose a value of $|\Delta_t|$ which is quite a bit larger than the natural linewidth $\Gamma_e= 2\pi\times 6$ MHz of state $|e_L\rangle$ because we instead aim for $\Delta OD=0$ combined with a large value of $\Delta\beta$. (Appendix)

Zeroing the conditional optical depth $\Delta OD$ is important for achieving a high post-selected fidelity of the gate. To see this, assume that we attempt to create the entangled state $|\psi_i\rangle= (|LH\rangle-i|RV\rangle)/\sqrt2$ but instead obtain the normalized output state
\begin{multline}
|\psi_\xi\rangle
= \frac12 e^{-\xi_0}(|RR\rangle + e^{-\xi_2}|RL\rangle
\\
+e^{-\xi_1}|LR\rangle+ e^{-\xi_1-\xi_2-\xi_3} |LL\rangle)
+ c_\text{abs}|\psi_\text{abs}\rangle
.\end{multline}
Here the normalized state vector $|\psi_\text{abs}\rangle$ with a complex amplitude $c_\text{abs}$ contains all those components of the state that correspond to absorption of at least one photon. Hence, $|\psi_\text{abs}\rangle$ is orthogonal to the states $|RR\rangle$, $|RL\rangle$, $|LR\rangle$, and $|LL\rangle$. When finally post-selecting upon detection of one control and one target photon, then all other properties of $|\psi_\text{abs}\rangle$ become irrelevant, so it is unnecessary to detail them here. $1-|c_\text{abs}|^2$ is the probability that the state contains two photons.

All other coefficients are expressed in terms of the complex numbers $\xi_i$. A post-selected fidelity of unity would be achieved for $\xi_1= 0$ and $\xi_2= \xi_3= i\pi$ along with arbitrary $\xi_0$. We decompose the $\xi_i$ as $\text{Re}(\xi_i)= OD_i/2$ and $\text{Im}(\xi_i)= -\beta_i$. The resulting phase shifts $\beta_i$ are obviously related to the phase shifts discussed above, namely $\beta_1= \beta_{c,L}-\beta_{c,R}$ and $\beta_2= \beta_L-\beta_R$ are the differential phase shifts of the control and target qubit, respectively, $\beta_3= \Delta\beta$ is the conditional phase shift, and $\beta_0= \beta_R+\beta_{c,R}$ is a global phase shift, which is of little relevance. The interpretation of the $OD_i$ is analogous, in particular $OD_3= \Delta OD$.

It is a crucial point that using linear optics, including polarisation-selective attenuators, it is experimentally easy to implement arbitrary single-qubit operations, as long as they maintain or attenuate intensities. If they act only on the first (second) qubit they will modify only $\xi_0$ and $\xi_1$ ($\xi_0$ and $\xi_2$). However, single-qubit operations can never modify $\xi_3$. To modify $\xi_3$, one needs interactions between the two qubits, which in our experiment result from Rydberg blockade during the passage of the target photon through the atomic ensemble. Hence, any undesired value of $\xi_1$ or $\xi_2$ which might be created when the photons pass through the atomic ensemble can be compensated before or after that passage with easy-to-implement single-qubit operations. But possible deviations from $\Delta\beta= \pi$ or $\Delta OD= 0$ incurred during passage through the atomic ensemble cannot be compensated before or after that passage. This is why we aim to achieve $\Delta\beta= \pi$ and $\Delta OD= 0$ simultaneously upon passage through the atomic ensemble.

In our experiment, the single-qubit operations needed to compensate $OD_1\neq0$ and $OD_2\neq0$ are easily implemented as follows. For the control qubit, $OD_1\neq0$ is compensated by placing an appropriate neutral density filter (NDF) into one arm of the first interferometer. As $OD_1\neq OD_2$, we additionally temporally switch the radio-frequency (rf) power driving the AOM in the first interferometer to obtain the appropriate compensation for the target qubit. At fixed optical input power, this implementation would reduce the overall count rate, but in our experiment it does not because we can increase the input power. This is because to avoid a degradation of the post-selected fidelity, the mean photon number per pulse impinging onto the atomic ensemble must be below unity. Otherwise possible excess photons might become entangled with the other photons and if the excess photons happen not to be detected, the post-selected fidelity deteriorates. The first interferometer, however, contains only linear optical elements. They cannot produce entanglement with the excess photons so that loss of possible excess photons in the first interferometer has no effect onto the post-selected fidelity.

\bigskip
\noindent
\textbf{Choice of Parameter Values}

\noindent
In the experiment, we need to choose a large number of parameters. A large principal quantum number has the advantage of yielding a large van der Waals coefficient $C_6$, which results in a large blockade radius $r_b$ which in turn makes $\Delta\beta$ large. On the other hand, a large principal quantum number tends to yield a large dephasing rate $\gamma_{rg}$, which is undesirable. This motivates the choice of the F\"orster resonance between the Rydberg states $67S$ and $69S$ in $^{87}$Rb, which increases $|C_6|$ by a factor of $\sim60$ compared to $69S$ and $69S$ \cite{Tiarks:14}.

In the presence of a nonzero dephasing rate $\gamma_{rg}$, the parameter regime in which $\Delta\beta=\pi$ and $\Delta OD=0$ can be achieved simultaneously is limited (Appendix). In particular, it is advantageous to choose a transition for the $|L\rangle$ polarisation of the target qubit, which has a large electric dipole matrix element. The largest electric dipole matrix element in an optical excitation from the $5S$ ground state is the one on the cycling transition $|g\rangle\leftrightarrow|e_L\rangle$. This is why we choose this transition for the target photon. Hence, the internal atomic state must initially be $|g\rangle$.

We want the frequency of the incoming control and target photons to be roughly the same, so that the control photon should also be near resonant with the $5S \leftrightarrow 5P$ transition. With the atomic population initially prepared in state $|g\rangle$, the $|L\rangle$ polarisation of the control photon must use the same transition as the target photon, because the other fine- and hyperfine-structure components of the $5P$ state do not have $m_F=3$ components. The disadvantage of this choice of transitions is that without selectively frequency shifting the $|R\rangle$ polarisation, one cannot easily achieve storage of the $|R\rangle$ polarisation in the $F=1$ ground state.

In principle, one could alternatively store the $|R\rangle$ polarisation in a Rydberg state with yet a different principal quantum number, e.g.\ $68S$. In terms of technical effort, this would require setting up a third 480 nm laser. This scenario would have the disadvantage of reducing $\Delta\beta$ because a stored $|R\rangle$ photon could also create Rydberg blockade for the target photon. As the $C_6$ coefficients will typically differ, $\Delta\beta$ would not vanish completely, but a substantial reduction in $\Delta\beta$ could already be a significant problem. Another alternative option could be to delay the $|R\rangle$ polarisation in an optical fibre. This would require an active stabilization of the optical path length difference, resulting in a technical effort similar to the stabilization of the modified Mach-Zehnder interferometers. The two interferometers used in our setup offer the advantage that the optical path length which must be stabilized does not contain the ultracold atomic gas, which could otherwise make the stabilization technically more challenging. Another advantage of the two interferometers is that the rf power of the AOM can be switched to compensate $OD_1\neq OD_2$, as discussed above.

We typically choose an atomic peak density of $\varrho= 2\times 10^{12}$ cm$^{-3}$ and a length of the medium of $L= 60$ $\mu$m, similar to Ref.\ \cite{Tiarks:16}. The choice of $\varrho$ reflects a compromise. On the one hand, large $\varrho$ is preferable because it increases $\Delta\beta$ and storage efficiency. On the other hand, small $\varrho$ reduces the dephasing rate $\gamma_{rg}$ \cite{Baur:14}. The choice of $L$ also reflects a compromise. On the one hand, large $L$ is preferable because it increases storage efficiency and target photon visibility (Appendix). On the other hand, small $L$ reduces absorption loss during propagation. The dephasing rate $\gamma_{rg}$ describes a decay of coherence between states $|g\rangle$ and $|r\rangle= |67S\rangle$. For the density and principal quantum number used here, we typically measure $\gamma_{rg}= 1$ $\mu$s$^{-1}$ (Appendix). With these parameters, a detailed model (Appendix) can be used to calculate optimal values for the coupling Rabi frequency $\Omega_{L,t}$ and the signal detuning $\Delta_t/2\pi$, which we implement to a good approximation in our experiment.

With these choices, we carefully zero $\Delta OD$ by fine tuning the target two-photon detuning to be slightly off from the two-photon resonance (Appendix). In addition, we carefully fine tune the combination of atom number and length of the medium to achieve $\Delta\beta=\pi$ and $\beta_2= \pi$. The latter choice is motivated in the Appendix.

\bigskip
\noindent
\textbf{Acknowledgements}
This work was supported by Deutsche Forschungsgemeinschaft through Nanosystems Initiative Munich.

\appendix

\section{Experimental Aspects}

\subsection{Dipole Trap and Atomic Ensemble}

\label{sec-dipole-trap}

The atomic gas is trapped in an optical dipole trap, similar to Ref.\ \cite{Tiarks:16}. A light beam at a wavelength of 1064 nm propagates horizontally along the $z§$ axis and provides radial confinement with a measured radial trapping frequency of 100 Hz and negligible axial confinement with an estimated trapping frequency of typically 0.1 Hz. With a temperature of typically 0.6 $\mu$K, the atomic ensemble is cigar shaped with a Gaussian density profile in the radial direction with an estimated root-mean-square (rms) radius of $\sigma_r= 12$ $\mu$m. The EIT coupling light fields at 480 nm create repulsive potentials for the ground-state atoms. These potentials have negligible effect because they are pulsed on with a low duty cycle, as e.g.\ in Ref.\ \cite{Tiarks:16}. The atomic ensemble is located in a glass cell with the nearest glass surfaces at a distance of 15 mm from the atoms. No measures have been taken to control the electric field at the position of the atoms.

Two light sheets at a wavelength of 532 nm provide a box-like longitudinal potential. In contrast to Ref.\ \cite{Tiarks:16}, these light sheets have an elliptically shaped spot with waists (1/$e^2$ radii of intensity) of 15 $\mu$m along the horizontal $z$ axis and 43 $\mu$m vertically and are operated at a power of 0.3 W each. The centers of the plug beams are separated by typically $\Delta z=100$ $\mu$m. The resulting axial density profile is approximately homogeneous with an estimated full width at half maximum (FWHM) of typically $L=60$ $\mu$m. With an atom number of typically $1.3\times 10^5$, we estimate an atomic peak density of typically $\varrho= 2.4 \times 10^{12}$ cm$^{-3}$. A magnetic field of 14 $\mu$T is applied along the symmetry axis of the dipole trap to stabilise the orientation of the atomic spins.

\subsection{Signal and Coupling Light for EIT}

Rydberg EIT is created using the 780-nm signal beam (red in Fig.\ \ref{fig-setup}) together with Rydberg coupling light. One Rydberg coupling beam at 480 nm (blue in Fig.\ \ref{fig-setup}) counterpropagates the signal beam. It is used to store and retrieve the $|L\rangle$ polarisation of the control pulse. The other Rydberg coupling beam at 480 nm (also blue in Fig.\ \ref{fig-setup}) copropagates with the signal beam. It is used to create detuned Rydberg EIT for the $|L\rangle$ polarisation of the target pulse. Both Rydberg coupling beams are overlapped with and separated from the EIT signal beam using dichroic mirrors. The Rydberg coupling beams have powers of typically $P_{L,c}= 140$ mW and $P_{L,t}= 26$ mW. Their waists of $w_{L,c}=21$ $\mu$m and $w_{L,t}=12$ $\mu$m are both large compared to the waist of the EIT signal beam of $w_s=8$ $\mu$m so that spatial inhomogeneities in the coupling intensities sampled by the signal light are no major concern. From these parameters, we estimate coupling Rabi frequencies of $\Omega_{L,c}/2\pi= 25$ MHz  and $\Omega_{L,t}/2\pi= 20$ MHz.

The ground-state coupling light beam (cyan in Fig.\ \ref{fig-setup}) required to store and retrieve the $|R\rangle$ polarisation of the control photon propagates perpendicularly to the EIT signal light. It has a waist of $w_g=64$ $\mu$m and a power $P_g$ between 0.8 and 1.6 $\mu$W, from which we estimate a coupling Rabi frequency $\Omega_g/2\pi$ between 6 and 8 MHz.

The 780-nm EIT signal light for the control and target pulses comes from the same laser. It is split into two beams to individually manipulate polarisation, timing, frequency, and power of the control and target signal input pulses. The two beams are recombined on a 50:50 non-polarising beam splitter (NPBS). Whenever we quote the input polarisation of a photon, we refer to the polarisation at the point after this recombination, which is just before entering the first interferometer shown in Fig.\ \ref{fig-setup}. As the EIT signal light pulses are created from a laser, they have Poissonian photon number statistics. Typical values for the average photon numbers impinging onto the atomic ensemble are 0.33 and 0.50 for the control and target pulse, respectively.

The incoming control signal pulse has the temporal shape of a Gaussian that is cut off in the center. Without cutting, its rms width would be 0.2 $\mu$s. The accompanying control coupling light fields at 480 nm and 780 nm are on for 1 $\mu$s during storage and for 2 $\mu$s during retrieval.

The incoming target signal pulse has a rectangular pulse shape and is 2.6 $\mu$s long. The accompanying target coupling light at 480 nm is on for 4.1 $\mu$s to leave room for the $\sim$0.4 $\mu$s EIT group delay experienced by the target polarisation $|L\rangle$. As the target polarisation $|R\rangle$ does not experience EIT, it acquires negligible group delay. Hence, the longitudinal wave-packet overlap of the target polarisations $|R\rangle$ and $|L\rangle$ deteriorates during propagation through the medium. Hence, in our data analysis of the transmitted target pulse, we include only those typically 2.0 $\mu$s in which the target polarisations $|R\rangle$ and $|L\rangle$ overlap well. This reduces the count rate. The coherence time of the incoming target photon is much longer than the EIT group delay so that temporal coherence is not a concern in this differential delay of wave packets.

\subsection{Zeroing the Conditional Optical Depth}

\begin{figure}[!t]
\centering
\includegraphics[scale=1.2]{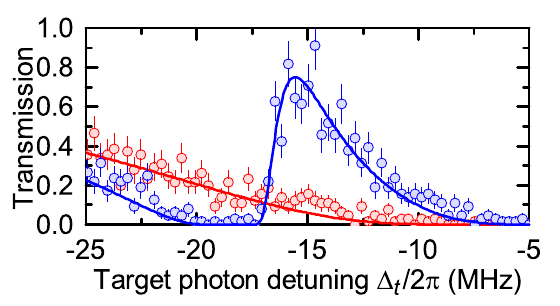}
\caption{
Target-light Rydberg EIT used for zeroing the conditional optical depth. The transmission of an $|L\rangle$ polarised target photon (blue) exhibits an EIT peak, with a maximum at the two-photon resonance at $\Delta_t/2\pi\sim-15$ MHz. This feature is absent if the target coupling light is switched off (red). The lines show fits of Eq.\ \eqref{chi} to the data. They cross at $\Delta_t/2\pi\sim-17$ MHz. Here, the conditional optical depth $\Delta OD$ is approximately zero.}
\label{fig-zero-OD}
\end{figure}

As discussed in the Methods Section, zeroing the conditional optical depth $\Delta OD$ is important for achieving a high post-selected fidelity of the gate. At a given value of the target coupling detuning, we use the target signal detuning $\Delta_t$ to tune $\Delta OD$ to zero. Fine tuning $\Delta_t$ based on data post-selected upon detection of one control and one target photon would be quite time consuming. As an approximate but much faster method, we consider Fig.\ \ref{fig-zero-OD}, which compares the target transmission with and without the target coupling laser, both recorded without sending a control photon into the system. Hence, no post-selection upon a retrieved control photon is needed and the data acquisition rate is much higher.

To understand, why Fig.\ \ref{fig-zero-OD} helps finding the two-photon detuning for which $\Delta OD=0$, we consider the step-function approximation discussed in section \ref{sec-blockade}. It suggest that in Fig.\ \ref{fig-zero-OD}, the blue and red data can be regarded as measures for $\Im\chi_u$ and $\Im\chi_b$, where $\chi_u$ and $\chi_b$ denote the susceptibility in the unblocked and blocked volume, respectively. In this approximation, $\Delta OD=0$ is obtained for that value of $\Delta_t$ where the two lines in Fig.\ \ref{fig-zero-OD} cross.

\subsection{Quantum Memory}

\label{sec-memory}

The retrieval efficiency from the Rydberg state exhibits damped oscillations as a function of the dark time \cite{Baur:phd, Mirgorodskiy:17}. For the parameters of our experiment, the first revival of the efficiency occurs at a dark time of 4.5 $\mu$s, which is why we choose this value for the dark time.

To characterise the quantum memory, we perform single-photon polarisation tomography of the retrieved light. The normalised single-photon Stokes parameters are $S_i= (P_i-P_{i^\perp})/(P_i+P_{i^\perp})$, where $i\in\{H,D,R\}$ denotes horizontal, diagonal (45$^\circ$), and righthanded circular polarisation, $P_i$ the power behind a polariser which transmits the polarisation $i$, and $H^\perp= V$, $D^\perp= A$, and $R^\perp= L$ denote vertical, anti-diagonal (-45$^\circ$), and lefthanded circular polarisation. The visibility, also known as the degree of (maximum) linear polarisation, is $V= (S_H^2+S_D^2)^{1/2}$. The azimuth $\varphi$ is defined modulo $2\pi$ by $S_H= V\cos\varphi$ and $S_D= V\sin\varphi$. For $S_R= 0$, $V$ is a measure for how much coherence there is between the $|R\rangle$ and $|L\rangle$ polarisations. We denote the visibilities of the control and target qubits as $V_c$ and $V_t$, respectively.

\begin{figure}[!t]
\centering
\includegraphics{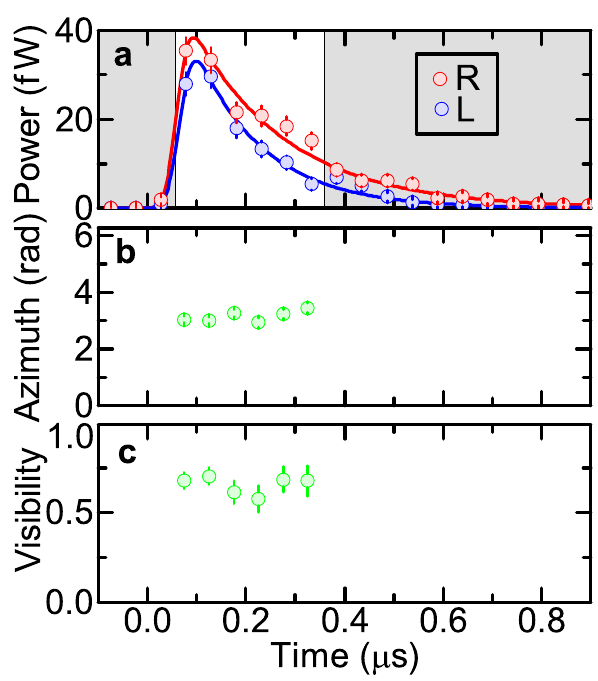}
\caption{
Performance of the quantum memory for the control qubit. \textbf{a} The retrieved power is well balanced between $|R\rangle$ and $|L\rangle$ at all times. \textbf{b}, \textbf{c} Azimuth and visibility of the retrieved light depend hardly on time. The visibility is high.
}
\label{fig-memory}
\end{figure}

For an $|R\rangle$ polarised input, the retrieved light is found to have a fraction $\epsilon_R= 4.8(5)$\% of its energy in $|L\rangle$. Likewise, for an $|L\rangle$ polarised input we obtain $\epsilon_L= 2.5(6)$\% in $|R\rangle$. Hence, the two polarisations $|R\rangle$ and $|L\rangle$ are well maintained during storage. To test how well coherence between $|R\rangle$ and $|L\rangle$ is maintained, we choose an input superposition such that the $|R\rangle$ and $|L\rangle$ components of the output energy are fairly well balanced. The results of this measurement are shown in Fig.\ \ref{fig-memory}. Fig.\ \ref{fig-memory}a shows that the retrieved light powers in $|R\rangle$ and $|L\rangle$ are fairly well balanced at all relevant times. To minimise the influence of detector dark counts, we analyse data only in the unshaded region. Polarisation tomography yields the azimuth $\varphi$ and the visibility $V_c$. These quantities, shown in Figs.\ \ref{fig-memory}b and c, are to a good approximation time independent, which is advantageous because it implies that we can effortlessly average data over the full data analysis time window. This averaging yields $V_c= 0.66(2)$. The average fidelity of the quantum memory is $F_m= (2+2V_c+(1+\epsilon_R)^{-1}+(1+\epsilon_L)^{-1})/6= 0.875(7)$, post-selected upon detection of the target photon. For comparison, if light is sent through both interferometers in the absence of atoms, we measure $V= 0.97$ for an appropriate input state. Hence, $V_c$ is predominantly limited by the atomic memory, not by the interferometers. The measured value of the post-selected average fidelity clearly exceeds the classical limit of 2/3.

Various effects can limit the visibility $V_c$ achievable in the control qubit after retrieval from the ground and Rydberg state. This is because the phase of the retrieved light depends on the relative phase accumulated during the dark time between an atomic superposition and a local oscillator, somewhat like in Ramsey spectroscopy. But in contrast to normal Ramsey spectroscopy, here three local oscillators instead of one contribute, namely the ground-state coupling laser at 780 nm, the control Rydberg coupling laser at 480 nm, and the rf reference oscillator, which drives the PLLs in both interferometers. The large wavelength difference between these two lasers makes it difficult to achieve low relative phase noise between them. The very different principal quantum numbers of the atomic states result in large differential electric polarisabilities for static or dynamic electric fields. Overall, various effects contribute to phase fluctuations, e.g.\ phase fluctuations of both control coupling lasers, fluctuating Zeeman and Stark effects in the presence of fluctuating magnetic and electric fields, fluctuating differential light shifts in the presence of power fluctuations in the dipole trapping light and in the target coupling light, and a distribution of differential light shifts in the dipole trapping light and in the target coupling light resulting from the thermal position distribution of the atoms. Clearly, the large number of possible issues listed here offers room for future improvements.

Reaching well-balanced powers of the $|R\rangle$ and $|L\rangle$ polarisations in the retrieved light at all relevant times is nontrivial. First, the time scales at which the retrieved powers decay in Fig.\ \ref{fig-memory}a must be made independent of the polarisation. We achieve this by adjusting the ratio of the intensities of the two control coupling light fields. Second, the overall retrieved energies, i.e.\ the areas under the curves in Fig.\ \ref{fig-memory}a, must be made independent of the polarisation. We reach that by adjusting the input polarisation. A possible differential group delay between the retrieved $|R\rangle$ and $|L\rangle$ polarisations could, in principle, be removed by delaying one of the retrieval coupling pulses, but this is not needed in our experiment.

Likewise, reaching a time-independent value of the azimuth in Fig.\ \ref{fig-memory}b is nontrivial. Under general conditions we typically observe some linear slope in Fig.\ \ref{fig-memory}b. However, after tuning the frequency of the control coupling laser to achieve $\beta_{c,L}-\beta_{c,R}= 0$ for all dark times, as discussed in the Methods Section, the linear slope in Fig.\ \ref{fig-memory}b vanishes automatically.

\subsection{Count Rates}

The combined efficiencies for storage and retrieval of the control qubit for a dark time of $t_d=4.5$ $\mu$s in the presence of a target pulse are roughly $\eta_R\sim 10\%$ and $\eta_L\sim 3\%$ for the $|R\rangle$ and $|L\rangle$ polarisations, respectively. The transmission of an $|R\rangle$ polarised target qubit is $T_R= e^{-OD_R}\sim 77\%$. The transmission of an $|L\rangle$ polarised target qubit is roughly $T_L\sim 15\%$, in reasonable agreement with the transmission at the point, where the two lines in Fig.\ \ref{fig-zero-OD} cross. This yields an efficiency $\eta_iT_j$ between 8\% and 0.5\% that a photon pair impinging onto the atomic ensemble is transmitted by the atomic ensemble, depending on the input polarisations $i,j\in\{R,L\}$. Note that future improvements regarding these numbers are possible.

Let $P_\mathrm{shot}$ denote the probability of a coincidence detection of one control and one target photon in a single experimental shot, which consists of storage of a control photon, propagation of a target photon, and retrieval of the control photon. In addition to the above-discussed efficiency $\eta_iT_j$ of the gate, several technical issues contribute to $P_\mathrm{shot}$, namely the quantum efficiency of 0.5 of each avalanche photodiode (APD) used for detection, the probability that an incoming pulse contains zero photons in a given shot, the non-unity transmission through the second interferometer, and the non-unity transmission through two single-mode optical fibres, one of which is located right after the atomic ensemble, the other right after the second interferometer. However, photon loss in components in front of the atomic ensemble is easily compensated by increasing the input power, as discussed in the Methods Section.

In daily alignment, we typically measure $P_\mathrm{shot}= 1.3\times10^{-5}$ for the entangling-gate operation. Such an experimental shot is repeated every 100 $\mu$s as in Ref.\ \cite{Tiarks:16}. The illumination with light causes spontaneous emissions which result in loss of atoms by spontaneous evaporation from the shallow dipole trap. To avoid a noticeable drop in atom number, we perform only $10^4$ experimental shots and then prepare a new atomic ensemble. A new atomic sample is prepared every 18 s so that in one minute, we detect an average number of 0.4 coincidences.

\section{Modelling Details}

\subsection{Rydberg EIT}

The linear electric susceptibility $\chi$ for Rydberg EIT in a ladder-type atomic level scheme with signal transition $|g\rangle\leftrightarrow |e\rangle$ and coupling transition $|e\rangle\leftrightarrow|r\rangle$ can be modelled as, see e.g.\ Ref.\ \cite{Tiarks:16}
\begin{align}
\label{chi}
\chi
= i\chi_0 \Gamma_e \left(\Gamma_e-2i\Delta_s+\frac{|\Omega_c|^2}{\gamma_{rg}-2i(\Delta_c+\Delta_s)}\right)^{-1}
,\end{align}
where $\Gamma_e=1/(26\text{ ns})$ is the population decay rate of the intermediate state $|e\rangle$, $\gamma_{rg}$ is the dephasing rate between states $|g\rangle$ and $|r\rangle$, $\Omega_c$ is the Rabi frequency on the coupling transition, $\Delta_s= \omega_s-\omega_{s,\text{res}}$ and $\Delta_c= \omega_c-\omega_{c,\text{res}}$ are the single-photon detunings of signal and coupling light, respectively, and $\chi_0= 2\varrho|d_{ge}|^2/\epsilon_0\hbar\Gamma_e$ is the value of $|\chi|$ for $\Omega_c= \Delta_s= 0$, where $\epsilon_0$ is the vacuum permittivity, $d_{ge}$ the electric dipole moment for the signal transition, and $\varrho$ the atomic density. Propagation through a homogeneous medium of length $L$ multiplies the complex amplitude $\bm E_0(z)$ of the electric field $\bm E(z,t)= \frac12 \bm E_0(z) e^{-i\omega_s(t-z/c)}+\text{c.c.}$ of the light by a factor $e^{-OD/2+i\beta}$ with the optical depth $OD= k_sL\Im(\chi)$ and the phase shift $\beta= k_sL\Re(\chi)/2$, where $k_s=\omega_s/c$ is the vacuum wave vector of the signal light and $c$ the vacuum speed of light. According to Eq.\ \eqref{chi}, the optical depth reaches its maximum $OD_\text{max}= k_sL \chi_0$ for $\Omega_c= \Delta_s= 0$.

The parameters of the atomic ensemble listed in section \ref{sec-dipole-trap} can be used to calculate $OD_\text{max}$. A simple calculation in the limit of vanishing waist of the signal beam $w_s\to0$ yields a value of typically $OD_\text{max}\sim 40$.

The fit of Eq.\ \eqref{chi} to the target-light EIT data (blue) in Fig.\ \ref{fig-zero-OD} yields a coupling Rabi frequency of $\Omega_{L,t}/2\pi=12.5(3)$ MHz for the target light, in reasonable agreement with the above estimate, and a dephasing rate of $\gamma_{rg}= 1.2(3)$ $\mu$s$^{-1}$, similar to the value obtained in Ref.\ \cite{Tiarks:16}. Dephasing is implemented in the model as a Markovian process described by $\partial_t \rho_{rg}= -\frac12\gamma_{rg}\rho_{rg}$, as in Ref.\ \cite{Fleischhauer:05}. Here, $\rho$ is the density matrix. In the absence of additional driving terms, this would produce an exponential temporal decay of $\rho_{rg}$. This makes it easy to implement the dephasing in an analytical calculation but it is not guaranteed to give an accurate description. For example, when applying this ansatz to describe the retrieval efficiency as a function of dark time, a simple model predicts an exponential decay of the efficiency with a $1/e$ time of $1/\gamma_{rg}$. As discussed in section \ref{sec-memory}, we observe damped oscillations instead. This shows that the Markovian ansatz does not capture all the relevant physical effects. From this perspective, $\gamma_{rg}$ represents a lowest-order approximation of the dephasing. Hence, when making large changes to other parameters, such as $\Omega_c$ or the atomic density $\varrho$, there is no guarantee that $\gamma_{rg}$ remains unchanged. On the other hand, $\gamma_{rg}$ is useful for modelling because it gives analytic curves which fit well to the transmission and phase shift as a function of $\Delta_s$, see e.g.\ Fig.\ \ref{fig-zero-OD} or Ref.\ \cite{Tiarks:16}.

\subsection{Rydberg Blockade}

\label{sec-blockade}

The presence of a stored control excitation in the Rydberg state $|69S\rangle$ creates a van der Waals energy $V(r)= -C_6/r^6$ for the propagating Rydberg polariton of the target polarisation $|L\rangle$, where $r$ is the distance between the two excitations and $C_6$ the van der Waals coefficient. For the combination of Rydberg states $|67S,F{=}m_F{=}2\rangle$ and $|69S,F{=}m_F{=}2\rangle$ in $^{87}$Rb, quantum defect theory yields $C_6= 2.3\times 10^{23}$ a.u.\ \cite{Tiarks:14}, where one atomic unit (a.u.) equals $9.573\times10^{-80}$ Jm$^6$. The van der Waals potential effectively modifies $\omega_{c,\text{res}}$ and, hence, the target coupling detuning according to $\Delta_c(r)= \Delta_{c,u}-V(r)/\hbar$, where $\Delta_{c,u}$ is the unblocked value. This creates a corresponding $r$ dependence of $\chi$ according to Eq.\ \eqref{chi}. Obviously $\Delta_c(r)$ diverges for $r\to 0$. $\chi$, however, approaches a finite value, which we call the completely blocked value $\chi_b= \lim_{r\to0}\chi(r)$. This value could alternatively be obtained by switching the EIT coupling light off. Hence, it equals the response of the two-level atom $|g\rangle\leftrightarrow |e\rangle$. We denote the unblocked value as $\chi_u= \lim_{r\to\infty}\chi(r)$.

Inspection of $\Re\chi(r)$ for the parameters of our experiment shows that $\Re\chi(r)$ is well approximated by a step function. We define the blockade radius $r_b$ as the value of $r$ at which $\Re \chi$ reaches the arithmetic mean of its unblocked value and its completely blocked value $\Re[\chi(r_b)]= \frac12\Re(\chi_b+\chi_u)$. The step-function approximation sets $\Re\chi= \Re\chi_u$ for $r>r_b$ and $\Re\chi= \Re\chi_b$ otherwise. An analogous argument applied to $\Im\chi(r)$ yields a blockade radius $r_{b,i}$ defined by $\Im[\chi(r_{b,i})]= \frac12\Im(\chi_b+\chi_u)$ and a step-function approximation $\Im\chi= \Im\chi_u$ for $r>r_{b,i}$ and $\Im\chi= \Im\chi_b$ otherwise. Typically $r_{b,i}\sim r_b$. To illustrate this, we temporarily consider the simple example of zero dephasing $\gamma_{rg}=0$, large single-photon detuning $|\Delta_s|\gg \Gamma_e$, and zero two-photon detuning $\Delta_{c,u}+\Delta_s=0$. Here, one obtains the analytic results $r_b= |4C_6\Delta_s/\hbar\Omega_c^2|^{1/6}$ (see also Ref.\ \cite{Firstenberg:13}) and $r_b/r_{b,i}= (1+\sqrt2)^{1/6}\sim 1.16$.

In our experiment, Rydberg blockade manifests itself in the conditional optical depth $\Delta OD$ and the conditional phase shift $\Delta\beta$. Suppressing fluctuations in $\Delta OD$ and $\Delta\beta$ is important for avoiding decay of coherence between the $|L\rangle$ and $|R\rangle$ polarisations of the target qubit in the presence of a stored Rydberg excitation. As the Rydberg blockade radius of $r_b=16$ $\mu$m (see section \ref{sec-parameters}) is much larger than the EIT signal beam waist of $w_s=8$ $\mu$m, the transverse degrees of freedom within the signal beam are irrelevant for the Rydberg blockade. This produces an effectively one-dimensional (1D) situation, in which fluctuations in $\Delta OD$ and $\Delta\beta$ otherwise caused by different relative transverse positions of the control and target photons are suppressed. In addition, the medium is to a good approximation homogeneous in the longitudinal direction and the transverse size of the medium $\sigma_r= 12$ $\mu$m is much larger than the $w_s=8$ $\mu$m EIT signal beam waist. This suppresses fluctuations in $\Delta OD$ and $\Delta\beta$ otherwise caused by storage of the control photon at positions with different atomic densities.

The conditional optical depth $\Delta OD$ and the conditional phase shift $\Delta\beta$ are defined as the difference between the values obtained in the presence and in the absence of a stored Rydberg excitation. We consider a 1D model, in which the signal beam propagates along the $z$ axis, assume that the medium is homogeneous and extends from $z=0$ to $z=L$, use the step-function approximation, denote the position of the stored control excitation as $z_s$, and introduce $L_b= \min(r_b,z_s)+\min(r_b,L-z_s)$ to denote the length of the blocked volume and a similar expression for $L_{b,i}$. This yields $\Delta OD= k_sL_{b,i}\Im(\chi_b-\chi_u)$, $\Delta\beta= k_sL_b\Re(\chi_b-\chi_u)/2$, and $L_b\leq\min(2r_b,L)$.

\subsection{Choice of Parameter Values}

\label{sec-parameters}

The following simple consideration yields a necessary condition for achieving $|\Delta\beta|= \pi$. From Eq.\ \eqref{chi} one can show that $|\Re(\chi)|\leq \chi_0/2$. This yields the upper bound $|\Delta\beta|\leq k_sL_b\chi_0/2= OD_b/2$, where we abbreviated the blocked optical depth $OD_b= OD_\text{max} L_b/L= k_sL_b\chi_0$. Hence, for reaching $|\Delta\beta|= \pi$ we obtain the necessary condition $OD_b\geq 2\pi$. Note that $|\Delta\beta|\leq OD_b/2$ is a tight bound because it can be reached, at least hypothetically, namely if and only if $\gamma_{rg}=0$, $|\Delta_s|=\Gamma_e/2$, and $\Delta_{c,u}+\Delta_s= |\Omega_c|^2/8\Delta_s$. It turns out that for these parameters $\Delta OD=0$, which is good. But at the same time $\Im\chi_b= \Im\chi_u= \chi_0/2$ so that $OD= OD_\text{max}/2$. Combination with $L_b\leq L$ shows that the transmission $T= e^{-OD}$ of the target intensity has the upper bound $T\leq e^{-OD_b/2}= e^{-\pi}\sim 0.04$. For $L=4r_b$, as in our experiment, $L_b\leq 2r_b= L/2$ even yields $T\leq e^{-OD_b}= e^{-2\pi}\sim 2\times 10^{-3}$.

The problem of this very low target transmission can be mitigated when using parameter values which do not reach the bound $|\Delta\beta|=OD_b/2$. Obviously, this comes at the cost of the need for larger $OD_b$. A simple estimate is obtained when assuming for simplicity that $\gamma_{rg}=0$. Hence, the condition $\Delta OD=0$ yields $\Delta_{c,u}+\Delta_s= |\Omega_c|^2/8\Delta_s$ and $\Delta\beta/OD_b= -2\Delta_s\Gamma_e/(4\Delta_s^2+\Gamma_e^2)$. Using $\Im\chi_b= \chi_0 \Gamma_e^2/(4\Delta_s^2+\Gamma_e^2)$, one obtains
$OD= k_sL\Im\chi_b=-\frac{\Gamma_e}{2\Delta_s} \Delta\beta \frac{L}{L_b}$. When increasing $|\Delta_s|$ at fixed $L$, $L_b$, and $\Delta\beta$, the optical depth decreases as $1/|\Delta_s|$. Hence, it is advantageous to work at larger $|\Delta_s|$, which according to the above relation for $\Delta\beta/OD_b$ requires larger $OD_b$ to keep $\Delta\beta$ fixed. In this model the optical depth vanishes for $|\Delta_s|\to\infty$. This is unrealistic because in this limit it becomes crucial that the dephasing rate $\gamma_{rg}$ is actually nonzero. A more elaborate model optimizing the target transmission while taking $\gamma_{rg}\neq0$ into account is presented in the rest of this section. For simplicity, this model assumes $L_b=2r_b$ although this is not necessarily the case, see section \ref{sec-V-t}.

In the Methods Section, we already motivated the choice of several parameters, namely the Rydberg states $|67S\rangle$ and $|69S\rangle$, the transition $|5S_{1/2},F{=}m_F{=}2\rangle \leftrightarrow |5P_{3/2},F{=}m_F{=}3\rangle$ for $|L\rangle$ polarised signal light, the atomic density $\varrho\sim 2\times10^{12}$ cm$^{-3}$, and the length of the medium $L\sim60$ $\mu$m. This choice of the atomic levels fixes the parameters $2\pi/k_s= 780.24$ nm, $\Gamma_e/2\pi= 6.07$ MHz, $d_{ge}= 2.54\times10^{-29}$ Cm, and $C_6= 2.3\times 10^{23}$ a.u. Empirically, we find $\gamma_{rg}\sim 1.2$ $\mu$s$^{-1}$ for these parameters, see e.g.\ Fig.\ \ref{fig-zero-OD}. In addition, the sign of $\Delta_s$ should be chosen such that $C_6 \Delta_s< 0$, as discussed e.g.\ in Ref.\ \cite{Tiarks:16}. With these parameters, a 1D model with homogeneous atomic density yields $OD_\text{max}= 35$.

We have three parameters left to choose, namely $\Delta_{c,u}$, $|\Delta_s|$, and $|\Omega_c|$. We wish to satisfy two constraints $\Delta\beta= \pi$ and $\Delta OD=0$. With three independent parameters and two constraints, there is one degree of freedom left to choose. We decide to use this degree of freedom to minimize the optical depth. As we use the step-function approximation, we have $\Delta OD= 2k_sr_{b,i}\Im(\chi_b-\chi_u)$. Hence, the constraint $\Delta OD=0$ implies $\Im(\chi_u)= \Im(\chi_b)$ so that minimizing $\Im(\chi_u)$ or minimizing $\Im(\chi_b)$ is equivalent. Therefore we do not need to make a choice which of the two we want to minimize. The constraint $\Delta OD=0$ also implies that $r_{b,i}$ disappears from the problem.

The constraint $\Delta OD=0$ can be solved analytically to yield $\Delta_{c,u}= -\Delta_s+ [|\Omega_c|^2\Gamma_e+\gamma_{rg}(-4\Delta_s^2+\Gamma_e^2)]/8\Gamma_e\Delta_s$. The remaining task is to minimize $\Im(\chi_b)$ with free parameters $|\Delta_s|$ and $|\Omega_c|$ under the constraint $\Delta\beta=\pi$. With some effort this problem can be solved analytically using the method of Lagrange multipliers, yielding $|\Omega_c|^2= 6\gamma_{rg}(4\Delta_s^2+\Gamma_e^2)/\Gamma_e$ and $\Delta_s= \frac12\Gamma_e(\zeta+\sqrt{\zeta^2-1})\sgn(-C_6)$, where $\sgn(x)= x/|x|$ and where we abbreviated a dimensionless parameter $\zeta= (2/7) \left|{C_6}/{\hbar\gamma_{rg}}\right|^{1/7} \left({3\chi_0k_s}/{\pi}\right)^{6/7}$. Obviously, a physical solution requires $\Delta_s$ to be real, i.e.\ $\zeta\geq 1$. This puts a nontrivial constraint at the parameters for which the gate can function at all. For the above-listed parameters of our experiment, we obtain $\zeta= 2.6$. As discussed below, reducing $\zeta$ would result in lower transmission, which is undesired. If $\zeta$ were reduced below unity, it would become impossible to simultaneously achieve $\Delta\beta=\pi$ and $\Delta OD=0$.

The model suggests that the optimal parameter settings should be $\Delta_s/2\pi= -15$ MHz, $|\Omega_c|/2\pi= 13$ MHz, and a small two-photon detuning $(\Delta_{c,u}+\Delta_s)/2\pi= -1.3$ MHz. We adopt the first two suggestions to a good approximation. The energy of the Rydberg state is not known very accurately in the experiment. Hence, we experimentally fine tune the two-photon detuning to achieve $\Delta OD=0$.

Moreover, this 1D model predicts $r_b= 16$ $\mu$m, $OD_b= 19$, and $\Im(\chi_b)= \Im(\chi_u)= \chi_0\Gamma_e/4\zeta|\Delta_s|= 2.6\times10^{-3}$. From the latter, the model predicts a transmission of $e^{-k_sL\Im(\chi_u)}= 0.26$ for the $|L\rangle$ polarisation of the target photon. The value of the transmission, measured where the two lines cross in Fig.\ \ref{fig-zero-OD}, is approximately half as large. This deviation comes about because Fig.\ \ref{fig-zero-OD} happened to be measured at $OD_\text{max}\sim 55$, which is quite a bit larger than the typical value normally used in our experiment.

For these values, the parameter $\zeta^2\sim 7$ is large compared to unity. Expanding the above solution in a power series for large $\zeta$ yields
\begin{align}
\lefteqn{
\Im(\chi_b)
= \frac{\chi_0}{4\zeta^2}[1+O(\zeta^{-2})]
}
\\ & \notag
= \frac{49}{16} \left(\frac{\pi}{3k_s}\right)^{12/7} \left(\frac{\hbar\gamma_{rg}}{|C_6|}\right)^{2/7} \left(\frac1{\chi_0}\right)^{5/7}[1+O(\zeta^{-2})]
.\end{align}
The first equality shows that maximizing $\zeta$ is crucial for achieving small $\Im(\chi_b)$. The second equality shows that if future improvements can reduce the dephasing rate $\gamma_{rg}$, then $\Im(\chi_b)$ can be reduced. In this model, the hypothetical limit $\gamma_{rg}\to0$ would result in $|\Delta_s|\to\infty$, $\Omega_c\to0$, and $\Im(\chi_b)\to0$, i.e.\ perfect transmission. The radiative lifetime of the Rydberg state, of course, sets a fundamental lower limit on $\gamma_{rg}$. For the $|67S\rangle$ state in a room-temperature environment, this limit is $\gamma_{rg}=1/(0.14\text{ ms})$ \cite{Tiarks:14}, a factor of $\sim$200 lower than the present best-fit value. There is clearly room for improvements.

This analysis reveals that the dephasing rate $\gamma_{rg}$ is a crucial performance-limiting element which implies that the laissez-faire attitude with which some proposals neglect dephasing should be reconsidered. On the other hand, our success in demonstrating a gate with our comparatively simple scheme despite the presence of dephasing is clearly an encouraging step for tackling some of the more elaborate proposals for Rydberg-based photon-photon gates. For example, cavity-based proposals \cite{Hao:15,Das:16,Wade:16} offer a way to operate at lower atomic density and therefore hopefully lower dephasing without incurring a loss in the multi-pass blocked optical depth. In addition, they can operate on single-photon resonance and use the cavity mirrors to convert conditional absorption into a conditional phase shift. Both aspects promise large margins for improving the performance of the gate.

\subsection{Visibility of the Target Photon}

\label{sec-V-t}

Based on the finite length $L$ of the medium, we will now derive an upper bound on the target-photon visibility $V_t$ achievable in the present scheme. This limit comes about because in our experiment $L>r_b$ so that the length of the blocked volume $L_b$ depends on the position $z_s$ at which the control excitation is stored. As $z_s$ is a random variable, so are $L_b$ and $\Delta\beta$.

We assume that the quantum state of the stored excitation is initially $|u\rangle= \int dz u(z) |z\rangle$ with $\langle u|u\rangle=1$, where $u(z)= \langle z|u\rangle$ is the spatial wave function. After a target pulse containing a single photon in the input polarisation state $|H\rangle$ propagated through the medium, the two excitations are finally in the state $|\psi\rangle= \frac1{\sqrt2} \int dz (|R\rangle+e^{i\Delta\beta(z)}|L\rangle)\otimes u(z) |z\rangle$, in which the target-photon polarisations became entangled with the position $z$ of the stored excitation. As we eventually do not measure $z$, we take the partial trace over this degree of freedom, yielding a reduced density matrix for the target-photon polarisation $\rho_t= \frac12 [|R\rangle\langle R|+|L\rangle\langle L|+(V_te^{i\beta_4}|L\rangle\langle R|+\text{H.c.})]$ with parameters $\beta_4{\in}]{-}\pi,\pi]$ and $V_t\geq0$ defined by
\begin{align}
V_te^{i\beta_4}
= \int_{-\infty}^\infty dz |u(z)|^2e^{i\Delta\beta(z)}
.\end{align}
$V_t$ is the visibility introduced above. This result is plausible. Averaging $e^{i\Delta\beta(z)}$ over $z$ with weighting function $|u(z)|^2$ yields an effective phase factor with reduced magnitude.

For simplicity, we assume that $|u(z)|^2$ is constant inside the medium and zero outside the medium. A proper handling of the cases resulting from $L_b= \min(r_b,z_s)+\min(r_b,L-z_s)$ yields
\begin{widetext}
\begin{align}
\label{V-t}
V_te^{i\beta_4}
=
\begin{cases}
e^{i\Delta\beta_b}[1-\frac{2r_b}L-\frac{4ir_b}{L\Delta\beta_b} (1-e^{-i\Delta\beta_b/2})], & 2r_b<L \\
e^{i\Delta\beta_bL/2r_b} [ -1 +\frac{2r_b}L - \frac{4ir_b}{L\Delta\beta_b}(1-e^{i\Delta\beta_b (r_b-L)/2r_b})] ,& r_b<L<2r_b \\
e^{i\Delta\beta_bL/2r_b} ,& L<r_b,
\end{cases}
\end{align}
\end{widetext}
where $\Delta\beta_b= k_sr_b\Re(\chi_b-\chi_u)$ denotes the bulk value of $\Delta\beta$ that is obtained for $L\to \infty$ which implies $L_b=2r_b$. Obviously $V_t= 1$ for $L<r_b$ which is plausible because in this case the stored excitation always blocks the complete medium. In the opposite limit $L\to \infty$, one also obtains $V_t\to 1$ which is plausible because in this case $\Delta\beta(z_s)= \Delta\beta_b$ for almost all $z_s$.

\begin{figure}[!t]
\centering
\includegraphics[scale=1.2]{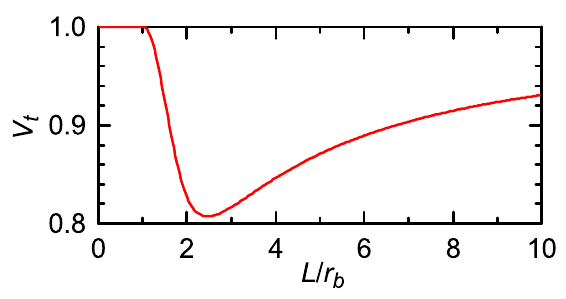}
\caption{Expected dependence of the visibility of the target pulse $V_t$ on the length of the medium $L$ in units of the blockade radius $r_b$. For $L/r_b\leq 1$ the complete medium is blocked, resulting in perfect visibility. For larger $L$ the conditional phase shift depends on the random position $z_s$ at which the control excitation is stored. As $z_s$ is not measured, the visibility is reduced. Parameters in the model are chosen such that the conditional phase shift becomes $\pi$ after averaging over $z_s$.}
\label{fig-target-visibility}
\end{figure}

In our experiment, the measured value of the conditional phase shift is $\beta_4$ and we tune the experimental parameters such that this equals $\pi$. To model this for a given value of $L/r_b$, we numerically search the value of $\Delta\beta_b$ which yields $\beta_4= \pi$ in Eq.\ \eqref{V-t}. Numerical results for the resulting values of $V_t$ are shown in Fig.\ \ref{fig-target-visibility}.

For $L\sim 4r_b$, as in our experiment, Fig.\ \ref{fig-target-visibility} yields $V_t\sim 0.85$. As other experimental imperfections will tend to reduce the measured value of $V_t$, this value is to be regarded as an upper bound on $V_t$. In Ref.\ \cite{Tiarks:16} also with $L\sim 4r_b$, we measured $V_t=0.75(14)$ which is a bit worse than the upper bound estimated here.

From this perspective, it would seem desirable to work at $L= r_b$ because here this model predicts $V_t=1$. The problem with this idea is that if the atomic density is kept the same, then $\zeta$ would be approximately halved, which would on one hand drastically reduce the target light transmission and on the other hand bring $\zeta$ close to unity so that it would be difficult to stably produce $\Delta\beta= \pi$. To avoid these problems, one could approximately double the atomic density but that would increase dephasing issues which would harm efficiency and post-selected fidelity. However, if one found a way to reduce the dephasing rate, working at $L=r_b$ would become an appealing option for increasing $V_t$ in the future. In our present experiment we accept a moderate reduction in $V_t$ at this point to avoid the problems arising from increased dephasing.

Alternatively, one could work at much larger $L$, because $V_t\to1$ for $L\to\infty$. However, as seen in Fig.\ \ref{fig-target-visibility} a large increase in $L$ would be needed to make a big difference for $V_t$. With the present experimental imperfections, the resulting reduction in the efficiency would outweigh the benefits from the increase in $V_t$.

\subsection{Choice of $\beta_2$}

To motivate the choice $\beta_2= \pi$, recall that we switch the rf power driving the AOM in the first interferometer to compensate for $OD_1\neq OD_2$. In principle, we could also switch the phase of the same rf field to compensate for $\beta_1\neq \beta_2$. As discussed in the Methods Section, the choice to compensate $OD_1\neq OD_2$ before the photons interact with the atomic ensemble has the advantage that a reduction of count rate can be avoided. As compensating a phase shift does not introduce photon loss, this argument does not apply to the compensation of $\beta_1\neq \beta_2$. Hence, we can just as well compensate $\beta_1\neq \beta_2$ after the photons interacted with the atoms, e.g.\ by switching the phase of the rf field driving the AOM in the second interferometer. However, some technical effort would be required to make switching the phase of one of these rf fields compatible with the PLL used to stabilize that interferometer against drift. Instead, we choose an equivalent method which is technically simpler.

In this context, it comes in handy that for detection, the photons are first sent though a 50:50 NPBS. Each output beam of the NPBS is sent through a set of waveplates for basis selection and then through a PBS. There is a total of four output beams of the two PBSs. Each of them is sent onto an APD. Hence, we could set the waveplates in front of one (the other) PBS to give the required compensation for the control (target) photon. The disadvantage of this method is that the NPBS does not selectively send the photons into the desired paths. Hence, one would lose a factor of four in count rate with this method. In principle, one could overcome this by replacing the NPBS with a device, such as an AOM, which allows it to time-dependently steer the photons into the desired paths.

For $\beta_2-\beta_1=\pi$, however, one can stick with the NPBS and use identical waveplate settings in front of both PBSs. This is because for $\beta_2-\beta_1=\pi$ the Stokes parameters $S_H$ and $S_D$ of the target photon simply have opposite sign compared to the Stokes parameters of the control photon. This inversion can be compensated after data acquisition during data processing. For the parameters of our experiment, the value of $\beta_2-\beta_1$ happens to be close to $\pi$ anyway. By fine tuning an experimental parameter, we can easily make it equal $\pi$. As we choose parameters such that $\beta_1=0$, this yields $\beta_2= \pi$.

\subsection{Fidelity of the Entangling-Gate Operation}

Fluctuations and nonideal average values in the experimental parameters reduce the fidelity of the entangling-gate operation. Here, we present a simple model of these effects, which yields an upper bound for the fidelity of the entangling-gate operation given the measured visibilities of the control qubit $V_c$ and target qubit $V_t$. Of course, as soon as one manages to improve the visibilities, the upper bound will improve as well. For simplicity, we assume that nonideal behaviour comes entirely from fluctuations in the phases.

We assume that each realisation of the experiment yields an output state vector in which all of the states $|RR\rangle$, $|RL\rangle$, $|LR\rangle$, and $|LL\rangle$ have equal population. Hence, the output state vector has the general form
\begin{multline}
|\psi_\beta\rangle
= \frac12e^{i\beta_0}(|RR\rangle + e^{i\beta_2}|RL\rangle +e^{i\beta_1}|LR\rangle
\\
+ e^{i\beta_1+i\beta_2+i\beta_3} |LL\rangle)
,\end{multline}
with four real phases $\beta_0$, $\beta_1$, $\beta_2$, and $\beta_3$, the significance of which was discussed in the Methods Section. The input state $|HH\rangle$ for the entangling-gate operation equals $|\psi_\beta\rangle$ with $\beta_1=\beta_2=\beta_3=0$. The ideal output state $|\psi_i\rangle$ equals $|\psi_\beta\rangle$ with $\beta_1=0$ and $\beta_2=\beta_3=\pi$. The state $|\psi_\beta\rangle$ has fidelity $F_\beta= |\langle\psi_i|\psi_\beta\rangle|^2= [2+\cos\beta_1-\cos\beta_2+\cos(\beta_1+\beta_2+\beta_3) -\cos(\beta_1-\beta_2)-\cos(\beta_1+\beta_3)+\cos(\beta_2+\beta_3)]/8$ with the ideal state.

We use $\overline{\cdots}$ to denote the average over the fluctuations of the phases $\beta_1$, $\beta_2$, and $\beta_3$. Considering the control qubit in the absence of a target qubit, we obtain the output Stokes parameters $S_{H,\beta_1}= \cos\beta_1$, $S_{D,\beta_1}= \sin\beta_1$ and $S_{R,\beta_1}= 0$ for the input state vector $|H\rangle$ and a specific value of the random variable $\beta_1$. The ensemble averages are $S_H= \overline{S_{H,\beta_1}}= \overline{\cos\beta_1}$, $S_D= \overline{\sin\beta_1}$, and $S_R=0$ so that the visibility is $V_c= V_1= [(\overline{\sin\beta_1})^2+(\overline{\cos\beta_1})^2]^{1/2}$. The fidelity between the single-qubit input state $|H\rangle$ and the single-qubit output state $\rho_\mathrm{out}$ is $F_1= \langle H|\rho_\mathrm{out}|H\rangle= (1+S_H)/2$. The average value $\overline{\beta_1}$ can be varied in the experiment. Assuming that $V_1$ remains unchanged when doing so, we obtain the maximum $F_{1,m}= \frac{1+V_1}2$. Likewise, considering only the target qubit in the absence of the control qubit, we obtain $V_2= [(\overline{\sin\beta_2})^2+(\overline{\cos\beta_2})^2]^{1/2}$ and $F_{2,m}= (1+V_2)/2$. In analogy to these expressions, we formally define $V_3= [(\overline{\sin\beta_3})^2+(\overline{\cos\beta_3})^2]^{1/2}$ and $F_{3,m}= (1+V_3)/2$.

We now turn to the entangling-gate operation. For the input state vector $|HH\rangle$, our model yields the output density matrix $\rho= \overline{|\psi_\beta\rangle\langle\psi_\beta|}$ and the fidelity $F_e= \langle\psi_i|\rho|\psi_i\rangle= \overline{F_\beta}$. To evaluate this expression, we use trigonometric identities such as $\cos(\beta_1+\beta_2)= \cos\beta_1\cos\beta_2-\sin\beta_1\sin\beta_2$. For simplicity, we assume that the fluctuations in the variables $\beta_1$, $\beta_2$, and $\beta_3$ are uncorrelated. Hence, $\overline{\cos\beta_1\cos\beta_2}= \overline{\cos\beta_1}\ \ \overline{\cos\beta_2}$ etc. We assume that the average values $\overline{\beta_i}$ are chosen such that $F_e$ is maximised at fixed values of $V_1$, $V_2$, and $V_3$. We also assume that the probability distribution function of each $\beta_i$ is symmetric around its average value. Hence, $F_e$ is maximised if we choose $\overline{\beta_1}=0$ and $\overline{\beta_2}=\overline{\beta_3}=\pi$. This yields $\overline{\sin\beta_i}= 0$, $\overline{\cos\beta_1}= V_1$, $\overline{\cos\beta_2}= -V_2$, $\overline{\cos\beta_3}= -V_3$. A straightforward calculation now yields
\begin{align}
F_e
= \frac{1+V_1}{2} \; \frac{1+V_2}{2} \; \frac{1+V_3}{2} + \frac{1-V_3}{8}
.\end{align}
Using the same methods, we find that the visibility $V_t$ measured in Ref.\ \cite{Tiarks:16} in the presence of a stored $|L\rangle$ control excitation is $V_t= V_2V_3$. Given the measured value $V_t$ but with unknown $V_2$ and $V_3$, the expression for $F_e$ is maximised if we assume $V_2=1$ and $V_t=V_3$. As additional experimental imperfections tend to lower $F_e$ even further, we obtain the upper bound $F_e\leq (1+V_c)\linebreak[1](1+V_t)/4+\linebreak[1](1-V_t)/8$. Using the value $V_c=0.66(2)$ measured here and the value $V_t= 0.75(14)$ measured in Ref.\ \cite{Tiarks:16}, we obtain the upper bound $F_e\leq 0.76(6)$. Several effects might contribute to the fact that the measured value of $F_e$ does not reach this upper bound, e.g., events in which more than one control or more than one target photon impinge on the atomic sample (Methods), a possible reduction of the visibility of the control qubit resulting from the Rydberg interactions with the target photon, or imperfections in the experimental determination of the optimal mean values of the phases $\overline{\beta_j}$.

It is interesting to investigate whether the assumption that all imperfections are modelled as phase fluctuations has a large effect onto the resulting upper bound for $F_e$ at given measured values of $V_c$ and $V_t$. To address this question, we developed an analogous model that is based purely on population fluctuations. As long as all fluctuations are small, we find that the resulting upper bound is identical to the upper bound estimated from pure phase fluctuations. In the experiment both types of fluctuations exist. But as the investigation of both extreme cases --- pure phase fluctuations or pure population fluctuations --- predict identical upper bounds, it seems likely that a model taking both types of fluctuations into account simultaneously will predict a similar upper bound, as long as all fluctuations are small. This result is plausible because when taking only phase fluctuations into account, the amount of phase fluctuations needed to explain the observed values of $V_c$ and $V_t$ is larger than if both population and phase fluctuations are included in the model.

\subsection{Comparison with a Phase Shift Induced by Excitation Hopping}

A recent experiment \cite{Thompson:17} studied an alternative mechanism for creating a conditional phase shift of $\pi/2$. This prompts two questions. First, why does the mechanism studied there not occur in our experiment? Second, if we used that mechanism instead of ours would that improve the performance of the photon-photon gate?

To answer the first question, we note that the conditional phase shift in Ref.\ \cite{Thompson:17} is created as a result of excitation hopping between a propagating Rydberg polariton coupling to the $|100S\rangle$ state and a stored Rydberg excitation in the $|99P\rangle$ state. In principle, an analogous process could occur in our experiment with the states $|67S\rangle$ and $|69S\rangle$. But in practice, it is negligible for two reasons. First, both excitations in our experiment are in $S$ states which makes hopping a much slower second-order process. Second, the states used in our experiment exhibit predominantly van der Waals interaction, not excitation hopping. To quantify this, the van der Waals interaction at distance $r$ can be characterised by $V_b(r)= -C_6/r^6$ and the second-order excitation hopping by $V_{ex}(r)= -\chi_6/r^6$, see e.g.\ Ref.\ \cite{Thompson:17}. We use quantum defect theory to calculate $C_6/\chi_6= 29$ for the states used in our experiment, which shows that the van der Waals interaction dominates.

To answer the second question, we compare the performance of the two systems. We first note that our present work goes far beyond the scope of Ref.\ \cite{Thompson:17} because we demonstrate a quantum gate, not just a conditional phase shift. When comparing experiments which create a conditional phase shift, such as Refs.\ \cite{Tiarks:16} and \cite{Thompson:17}, the crucial figures of merit are conditional phase shift, visibility, and efficiency.

In terms of the size of the conditional phase shift Ref.\ \cite{Thompson:17} was already outperformed by our previous work \cite{Tiarks:16} where we demonstrated a conditional phase shift of $\pi$, not just $\pi/2$. In theory, any system creating a small conditional phase shift could be cascaded many times to eventually reach a $\pi$ phase shift but in practice this is hardly ever done because it tends to cause severe setbacks in terms of visibility and efficiency. As a consequence, a phase shift of $\pi$ has evolved into the standard benchmark because, first, creating an even larger phase shift is typically not helpful and, second, single-qubit unitaries provide an invertible map between a CNOT gate and a $\pi$ phase gate, thus offering a simple possibility to compare with the performance of implementations of CNOT gates in completely different physical systems.

The role of the visibility in comparing systems which create conditional phase shifts is similar to the role of the post-selected fidelity in comparing quantum gates. It is a really crucial parameter that characterizes the degree of phase coherence in the system. In Ref.\ \cite{Tiarks:16} we observed a visibility of 75(14)\%. Unfortunately, the visibility seems to be missing completely in Ref.\ \cite{Thompson:17}, making a comparison in terms of the visibility impossible.

Finally, the efficiency observed in Ref.\ \cite{Thompson:17} was $\eta= 0.06\times 0.82= 0.049$ for storage and retrieval of the first photon and $T= 0.56\times 0.77= 0.43$ for transmission of the second photon. To obtain a realistic comparison with our work, Ref.\ \cite{Thompson:17} would have to double its conditional phase shift. One could hypothesize that sending the second photon twice through the medium might double the conditional phase shift. Based on this one could optimistically extrapolate the total efficiency to be at best $0.82\times 0.75 \; T^2 \eta = 0.0056$. The factor $T^2$ expresses the transmission after passing twice through the medium. The factor 0.82 describes the reduction of the retrieval efficiency because of Rydberg interactions during the second passage of the second photon. The factor 0.75 estimates the reduction of the retrieval efficiency expected from thermal atomic motion at a temperature of 20 $\mu$K because the dark time $t_d= 1.4$ $\mu$s between storage and retrieval must be doubled. Taking only thermal motion into account, the retrieval efficiency is expected to decay according to (see e.g.\ Refs.\ \cite{Baur:phd,Jenkins:12}) $e^{-(t_d/\tau)^2}$ where $\tau= \frac{\lambda_{dB}}{\sqrt{2\pi}v_r}= 4.5$ $\mu$s is the $1/e$ time, $\lambda_{dB}$ the thermal de Broglie wavelength, and $v_r$ the relative speed between a ground-state atom and a Rydberg atom created by differential photon recoil during storage. Hence, doubling $t_d$ yields a factor $e^{-(2t_d/\tau)^2}/e^{-(t_d/\tau)^2}= 0.75$. Of course, it is unpredictable whether the system studied in Ref.\ \cite{Thompson:17} will be able to reach the above extrapolation or whether additional experimental complications will occur. Anyway, this optimistic extrapolation would suggest that Ref.\ \cite{Thompson:17} and our work would perform almost identically in terms of efficiency.


\end{document}